\documentclass[usegraphicx,usenatbib]{mn2e} 

\usepackage{url}
\usepackage[breaklinks=true]{hyperref}
\usepackage{twoopt}

\usepackage{natbib}
\bibpunct{(}{)}{;}{a}{}{,} 

\usepackage[utf8x]{inputenc}
\usepackage[normalem]{ulem}
\usepackage{siunitx}
\usepackage{amsmath, amssymb}
\usepackage[farskip=0pt]{subfig}

\usepackage{multirow,bigstrut,ctable}
\usepackage[normalem]{ulem}
\usepackage{rotating}
\usepackage[varg]{txfonts} 

\def \deg         {$^{\circ}$}
\def \arcmin      {$^\prime$}
\def \arcsec      {$^{\prime\prime}$}

\def \mujybeam    {$\rm \mu$Jy\,beam$^{-1}$}

\newcommand{\Msun}{\text{$\rm M_\odot$}}

\newcommand{\beam}[2]{{#1}\arcsec$\times${#2}\arcsec}

\newcommand{\arcmind}[2]{$#1'\,\hspace{-1.7mm}.\hspace{.1mm}#2$}

\def \target       {PSZ1 G096.89+24.17}

\title[A mass-luminosity relation for radio relics]
      {A new double radio relic in PSZ1 G096.89+24.17 and a radio relic mass-luminosity relation}

\author[F.~de~Gasperin et~al.]{F. de Gasperin$^{1}$, R. J. van Weeren$^{2}$, M. Br\"uggen$^{1}$, F. Vazza$^{1}$, A. Bonafede$^{1}$, H.T. Intema$^{3}$
\\
$^{1}$ Universit\"at Hamburg, Hamburger Sternwarte, Gojenbergsweg 112, D-21029, Hamburg, Germany\\
$^{2}$ Harvard-Smithsonian Center for Astrophysics, 60 Garden Street, Cambridge, MA 02138, USA\\
$^{3}$ National Radio Astronomy Observatory, 1003 Lopezville Road, Socorro, NM 87801-0387, USA}

\begin{document}

\date{}
\pagerange{\pageref{firstpage}--\pageref{lastpage}} \pubyear{2014}
\maketitle

\label{firstpage}

\begin{abstract}
Radio relics are diffuse synchrotron sources in galaxy clusters that are believed to trace large-scale shock waves. We have discovered a new double radio relic system in PSZ1 G096.89+24.17 ($z=0.3$) and have carried out a full-polarization radio observation using the Westerbork Synthesis Radio Telescope (WSRT) at 1.4 GHz. The observation revealed the presence of two relics located on the two diametrically opposite sides of the cluster and hints of a central radio halo. The linear sizes of the relics are $\sim 0.9$ and $\sim 1.4$ Mpc.\\ 
We carried out an analysis of all known double radio relics by using radio, X-ray and Sunyaev-Zeldovich (SZ) data. We find that the radio luminosity of double relics is a steep function of the cluster mass, with $L_R \propto M^{2.83\pm0.39}$. If we include single radio relics, this relation is maintained. This dependence has implications for the origin of magnetic fields at the relic's locations.
\end{abstract}


\section{Introduction}
\label{sec:introduction}

Giant radio relics are extended sources of radio emission found in galaxy cluster outskirts and are believed to trace shock waves generated during clusters mergers. Shock waves are sites for particle acceleration via the diffusive shock acceleration (DSA) mechanism \citep{Drury1983,Blandford1987,Jones1991}. Relics are expected to be preferably found in cluster outskirts because the kinetic energy dissipated in merger shocks peaks at a distance of about half of the virial radius from the cluster centre \citep{Vazza2012}.

Of a particular interest are those systems showing two radio relics, usually found on opposite sides of the cluster. These systems are believed to be the result of a binary merger where the merger axis is nearly perpendicular to the line of sight. Owing to the rather clear geometry of the system, double radio relics can constrain some important quantities as the mass ratio and the impact parameter of the merger \citep{Roettiger1999,vanWeeren2012}, as well as the efficiency of electron acceleration \citep{Bonafede2012a}.

\cite{VanWeeren2009} and \cite{Bonafede2012a} studied the correlation of the linear size of radio relics in \textit{double} systems against the radio power, the projected distance and the spectral index. They found that relics with high radio power have larger linear sizes and that there is a tendency of larger relics to be mostly located at larger distances from the cluster centre and of smaller relics to have steeper spectral indices. These tendencies can be explained by the fact that the larger shock waves occur mainly in lower-density and lower-temperature regions, and have hence larger Mach numbers \cite[i.e. a flatter spectra;][]{Skillman2008,Vazza2009}. However, the number of double radio relics discovered so far is limited to 15 \citep{Feretti2012,Bonafede2012a} and the statistical significance of the aforementioned correlations is barely sufficient to put loose constraints on models \citep{Bonafede2012a}.

Up to now only around 50 clusters hosting radio relics are known and these are located principally at moderate redshift \citep[$z \lesssim 0.3$;][]{Feretti2012}. With the advent of the Planck satellite, many new massive (up to $M_{500} = 1.6 \times 10^{15}$~\Msun) clusters with $z < 1$ were discovered \citep{PlanckCollaboration2013}. According to simulations, several hundred relics are expected to be discovered by novel radio telescopes \citep{Nuza2012}.

We performed a search for double radio relic systems in all newly discovered Planck clusters and also all the previously known clusters by visual inspecting multiple radio surveys (NVSS\footnote{NRAO VLA Sky Survey \citep{Condon1998}}, WENSS\footnote{Westerbork Northern Sky Survey, \citep{Rengelink1997}}, VLSS\footnote{VLA Low-frequency Sky Survey, \citep{Cohen2007}}) at the position of know galaxy clusters (1743 objects from the MCXC catalogue and 1227 from the Planck catalogue). Here we report the first outcome of this search which is the discovery of a double radio relic system in the \target{} cluster.

In Sec.~\ref{sec:observation} we present the WSRT observation and data reduction. In Sec.~\ref{sec:cluster} we present the results of these observations. In Sec.~\ref{sec:catalogue} we provide an up-to-date catalogue of all known double radio relics and we discuss the correlations between relics and hosting-clusters properties. Conclusions are in Sec.~\ref{sec:conclusions}. Throughout this paper we assume a $\Lambda$ cold dark matter ($\Lambda$CDM) cosmology with $H_0 = 71$ km s$^{-1}$ Mpc$^{-1}$, $\Omega_m$ = 0.27 and $\Omega_\Lambda$ = 0.73. At the redshift of \target{} ($z=0.3$) 1\arcsec{} corresponds to $\sim 4.421$ kpc. All images are in the J2000 coordinate system.

\section{Observations}
\label{sec:observation}

The cluster was observed with the WSRT on February 7, 2014, for about 12 h using the default 21~cm set-up. Due to telescope upgrades, only 8 antennas participated in the observation. A total bandwidth of 160~MHz was recorded, spread over eight spectral windows of 20~MHz in bandwidth. Each spectral window was further subdivided into 64 frequency channels. All four linear polarization products were recorded. The calibrators 3C286 and 3C147 were observed at the start and end of the main observing run, respectively.

The data were calibrated with CASA\footnote{version 4.2, \url{http://casa.nrao.edu}} using the flux scale given by \cite{Perley2012}. The first step consisted of the removal of time ranges affected by shadowing. We then performed an initial bandpass correction using 3C147. Radio frequency interference was removed using the AOFlagger \citep{Offringa2012}. This initial bandpass correction prevents flagging data affected by the strong bandpass roll-off. After flagging, we again calibrated the bandpass and subsequently obtained complex gain solutions for the two calibrator sources. The channel-dependent polarization leakage terms were calibrated using the unpolarized source 3C137. The polarization angles were set using 3C286, carrying out a single correction per frequency channel. We then transferred all calibration solutions to the target source. For the target field, three rounds of phase-only self-calibration were performed. The data was imaged taking the spectral index into account during the deconvolution \cite[i.e., {\tt nterms=3}][]{Rau2011}. We have made a high-resolution (${\rm robust}=0$) image and a low-resolution image (${\rm robust}=1$) to gain sensitivity towards the diffuse emission. The final images were corrected for the primary beam attenuation. In the case of the low-resolution image a dominant source $\sim30$\arcmin{} to the south caused small artifacts and has been peeled.

The final maps at two different resolutions are shown in Fig.~\ref{fig:flux}: \beam{15}{14} (rms: 23 \mujybeam, robust=0) and \beam{21}{20} (rms: 28 \mujybeam, robust=1).

\section{The double radio relic in \target}
\label{sec:cluster}

\begin{figure*}
\centering
\subfloat[High-resolution 1.4 GHz map]{\includegraphics[width=\columnwidth]{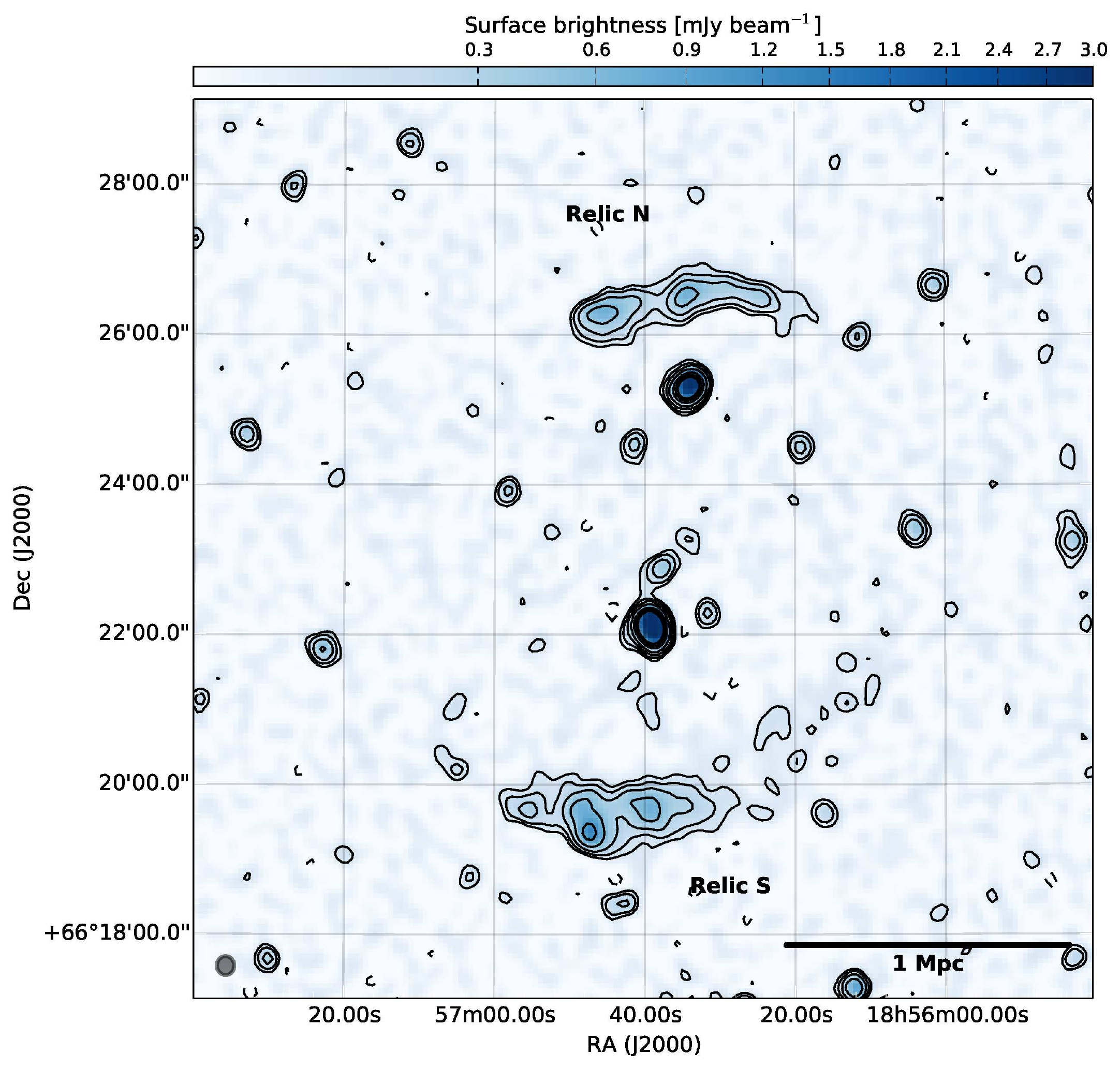}\label{fig:flux1}}
\subfloat[Low-resolution 1.4 GHz map]{\includegraphics[width=\columnwidth]{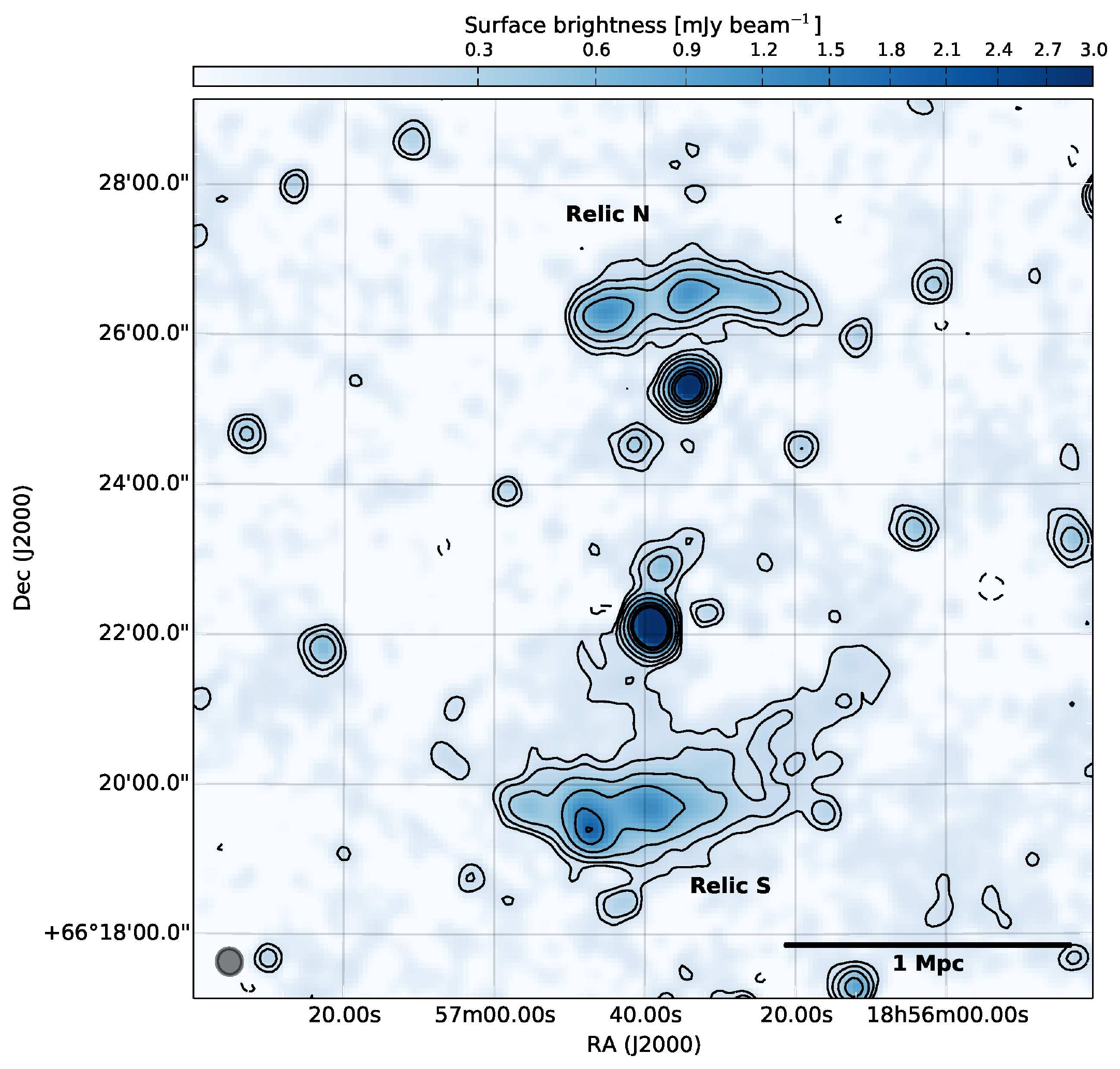}\label{fig:flux2}}
\caption{Left: WSRT 1.4 GHz image. High-resolution (beam: \beam{15}{14}) map made with robust=0 weighting. Contours are at the level of $(1, 2, 4, ...) \times\ 3\sigma_{\rm rms}$, $\sigma_{\rm rms} = 23$~\mujybeam. Dashed contours are at $-3\sigma$. Right: same map but at lower resolution (robust=1, beam: \beam{21}{20}) to enhance the extended emission. Contours are at the level of $(1, 2, 4, ...) \times\ 3\sigma_{\rm rms}$, $\sigma_{\rm rms} = 28$~\mujybeam. Dashed contours are at $-3\sigma$.}\label{fig:flux}
\end{figure*}

\target{} (also known as ZwCl 1856.8+6616) is a moderately massive galaxy cluster ($M_{500} = 4.40^{+0.45}_{-0.48} \times 10^{14}\ \Msun$) at $z=0.3$ \citep[spectroscopic measure by][]{PlanckCollaboration2013}. The ROSAT unabsorbed X-ray flux is $S_{\rm X} = 1.2\pm0.1 \times 10^{-12}$ erg s$^{-1}$ cm$^{-2}$ in the energy band: $0.1-2.4$ keV. At the redshift of 0.3 this provides an X-ray luminosity of $3.7 \times 10^{44}$ erg s$^{-1}$. Using the ($L$, $T$) relations provided by \cite{Pratt2009} we estimated a cluster temperature of T$\sim4$~keV.

\begin{figure}
\centering
\includegraphics[width=\columnwidth]{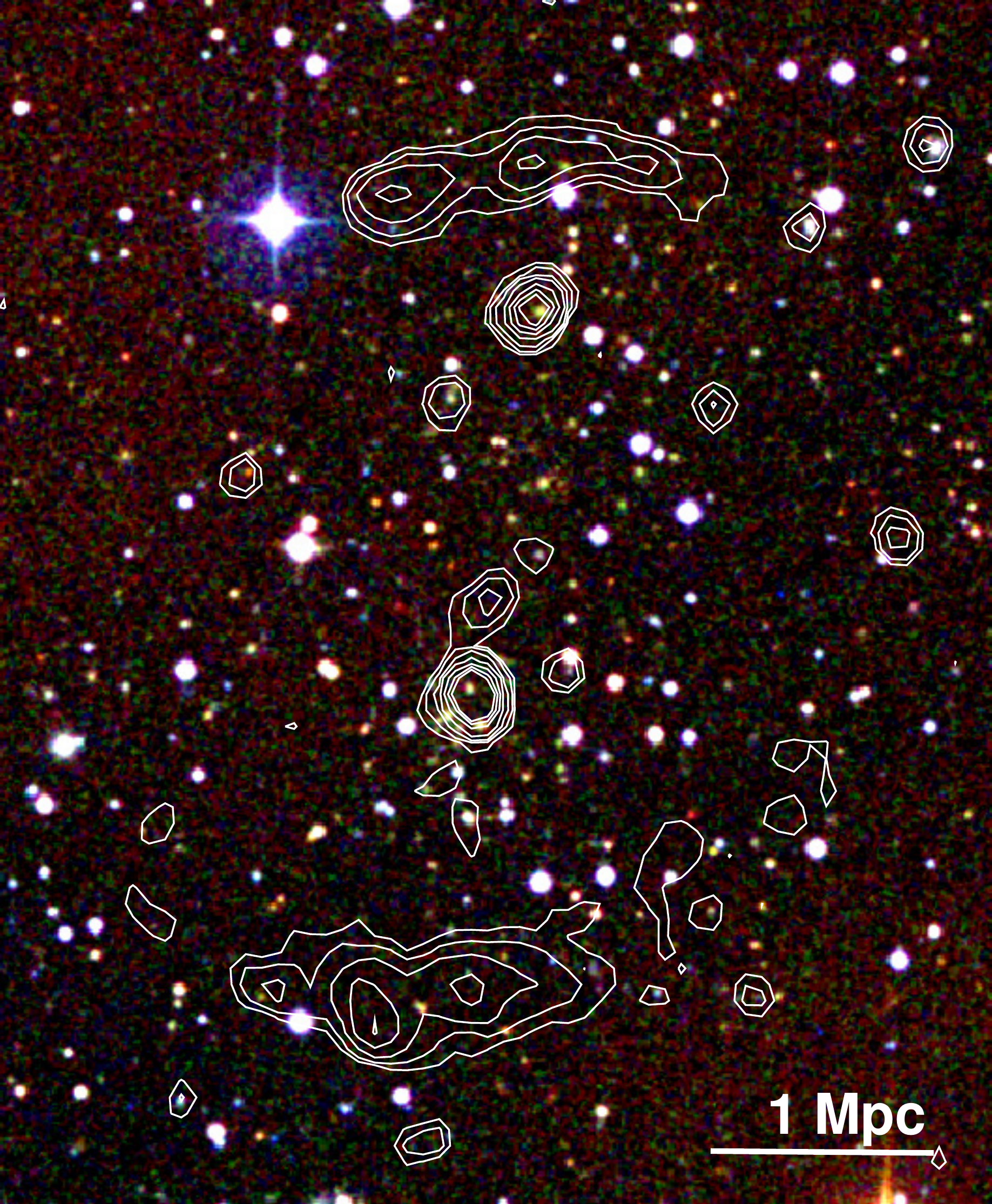}
\caption{DSS colour image (g, r and i band) of the cluster with overlaid the WSRT 1.4 GHz contours as in Fig.~\ref{fig:flux1}.}\label{fig:dss}
\end{figure}

\begin{figure}
\centering
\includegraphics[width=\columnwidth]{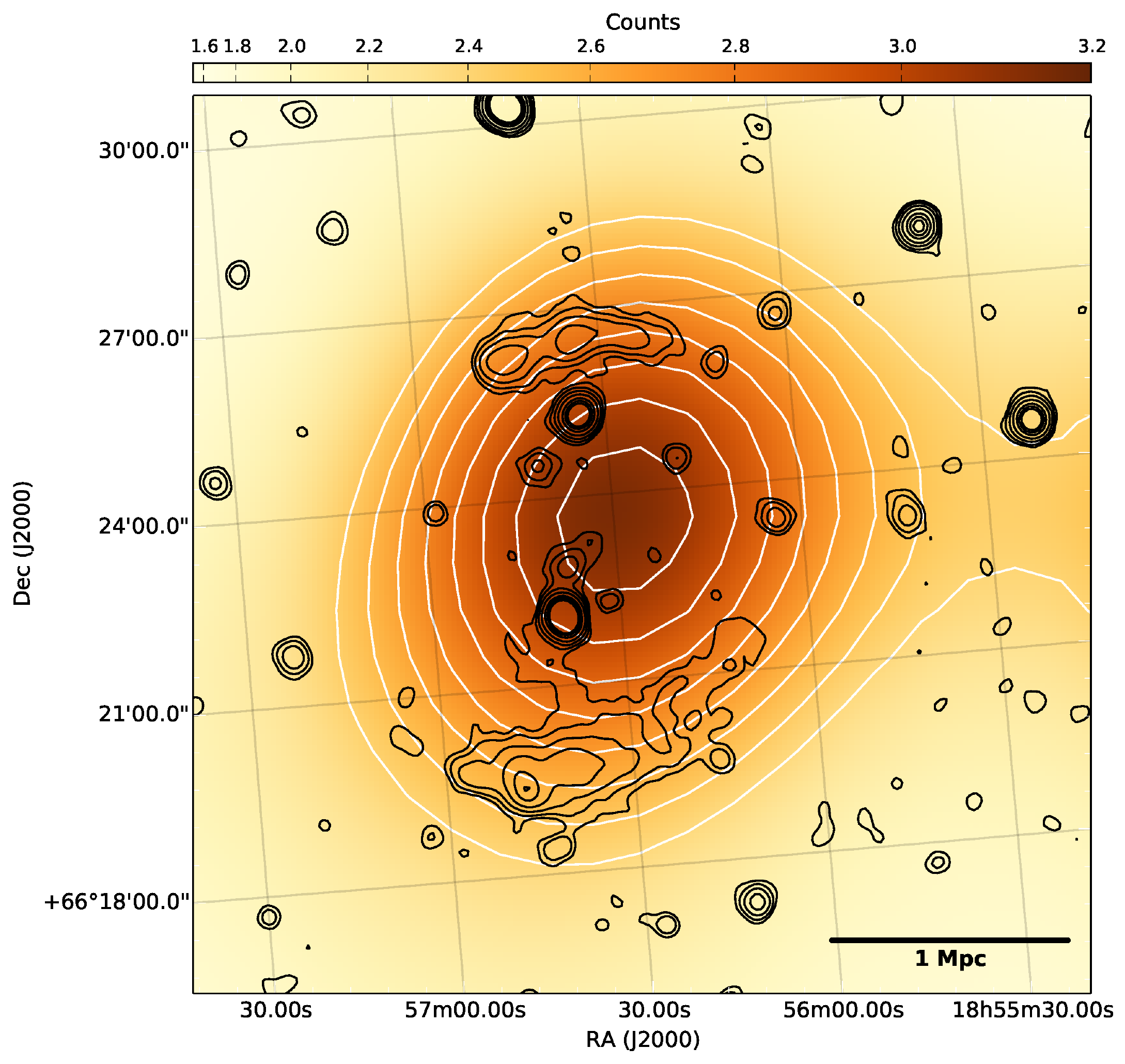}
\caption{Photon image in the broad band ($0.1-2.4$ keV) from the ROSAT all sky survey. No other X-ray observations of the cluster are currently available. Image has been smoothed with a two-dimensional Gaussian kernel of $\sigma=4$~pixels (1~$\textrm{pixel} = 45$~arcsec). White contours are at: $(2.4,2.5,...,3.3)$ counts. Black contours are from the WSRT 1.4 GHz observation as in Fig.~\ref{fig:flux2}.}\label{fig:rosat}
\end{figure}

The radio emission in the direction of this galaxy cluster is dominated by two strong point sources (one centrally located and one towards north) and by two arc-like extended structures labelled Relic N and Relic S in Fig.~\ref{fig:flux}. Relic N is about \arcmind{3}{4} long which is around 900 kpc at z=0.3, while relic S is around \arcmind{5}{5} long which translates into a linear size of $\sim 1.4$ Mpc. The complete extension of relic S is visible only in the low-resolution map (Fig.~\ref{fig:flux2}). A composite image from the Digitized Sky Survey (DSS) is shown in Fig.~\ref{fig:dss}. No obvious optical counterpart is visible in the location of the brightest spots of the extended radio emission. The two structures are at a distance of 770 kpc (Relic N) and 1145 kpc (Relic S) from the peak of the X-ray emission (see Fig.~\ref{fig:rosat}). The flux of the two extended sources has been extracted from the low-resolution map (Fig.~\ref{fig:flux2}) to ensure minimal flux is lost from the extended emission. Relic N has a flux density of $8.9\pm0.8$ mJy while Relic S has a flux density of $18.3\pm1.9$ mJy. Errors have been calculated by multiplying the map rms by the square root of the number of beams covering the source. We classify the two arc-like sources as radio relics because of (i) their morphology, (ii) their sizes, (iii) the absence of optical counterparts, and (iv) their location and orientation with respect to the cluster centre.

\begin{figure}
\centering
\includegraphics[width=\columnwidth]{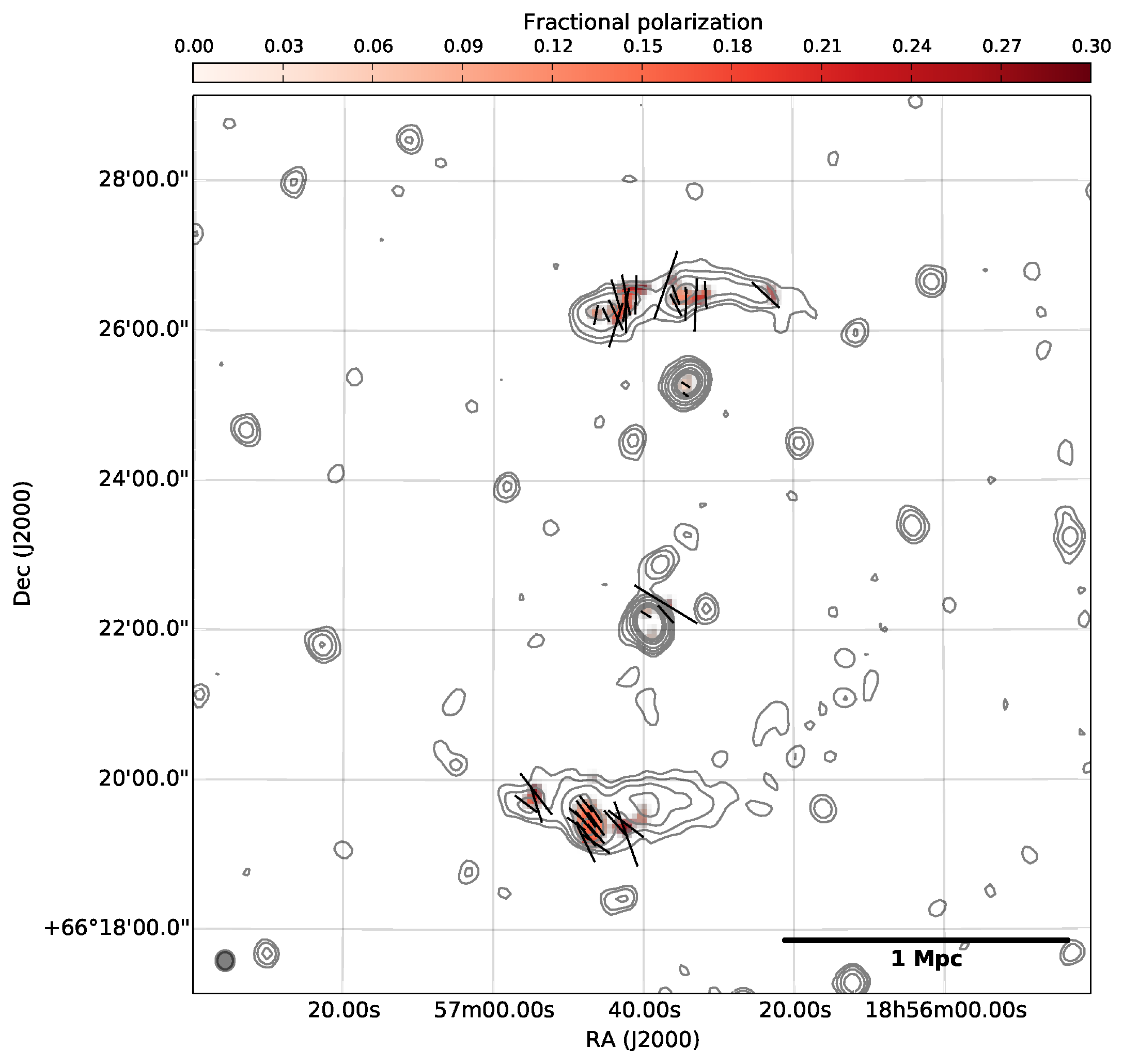}
\caption{Fractional polarization map with E-vectors displayed. The polarization angle has been corrected for galactic Faraday rotation. Black contours are the WSRT 1.4 GHz as in Fig.~\ref{fig:flux1}.}\label{fig:pol}
\end{figure}

Given the relic average surface brightness, the low signal-to-noise of the Q and U Stokes maps would allow a detection only for a fractional polarization $\gtrsim 10$\% and only in the brightest spots. A $\sim10-20$\% polarized emission is in fact detected on the brightest spots both in the northern and southern relics (see Fig.~\ref{fig:pol}). Using the map provided by \cite{Oppermann2012} we estimated a Faraday depth in the direction of \target{} of $\sim11.6$ rad m$^{-2}$. The polarization angle has therefore been corrected for galactic Faraday rotation by subtracting an angle of 31\deg. The northern relic appear to have E-vectors mostly aligned perpendicular to the relic extension. This is not the case for Relic S where the vectors are $\sim 45$ apart from being perpendicular to the relic extension.

Assuming that the magnetic field should be aligned with the relic extension as seen in other radio relics \citep[see e.g.,][]{VanWeeren2010a}, it would produce E-vectors perpendicular to the relic extension, as visible in Relic N. If we consider the merger axis to be slightly tilted, then we expect the ICM to produce a rotation in the polarization angle on the farther relic which should not be present in the closer one. The 45\deg misalignment with the polarization vector in the southern relic compared to the expected direction may be interpreted as a consequence of Faraday rotation. Polarization maps, with rotated polarization vectors in the southern relic compared to the expected orientation, together with the short distance between the relics and the peak of the cluster X-ray emission, suggest that the merge axis is slightly tilted compared to the plane of the sky, with the southern relic farther away from the observer.

\begin{figure}
\centering
\includegraphics[width=\columnwidth]{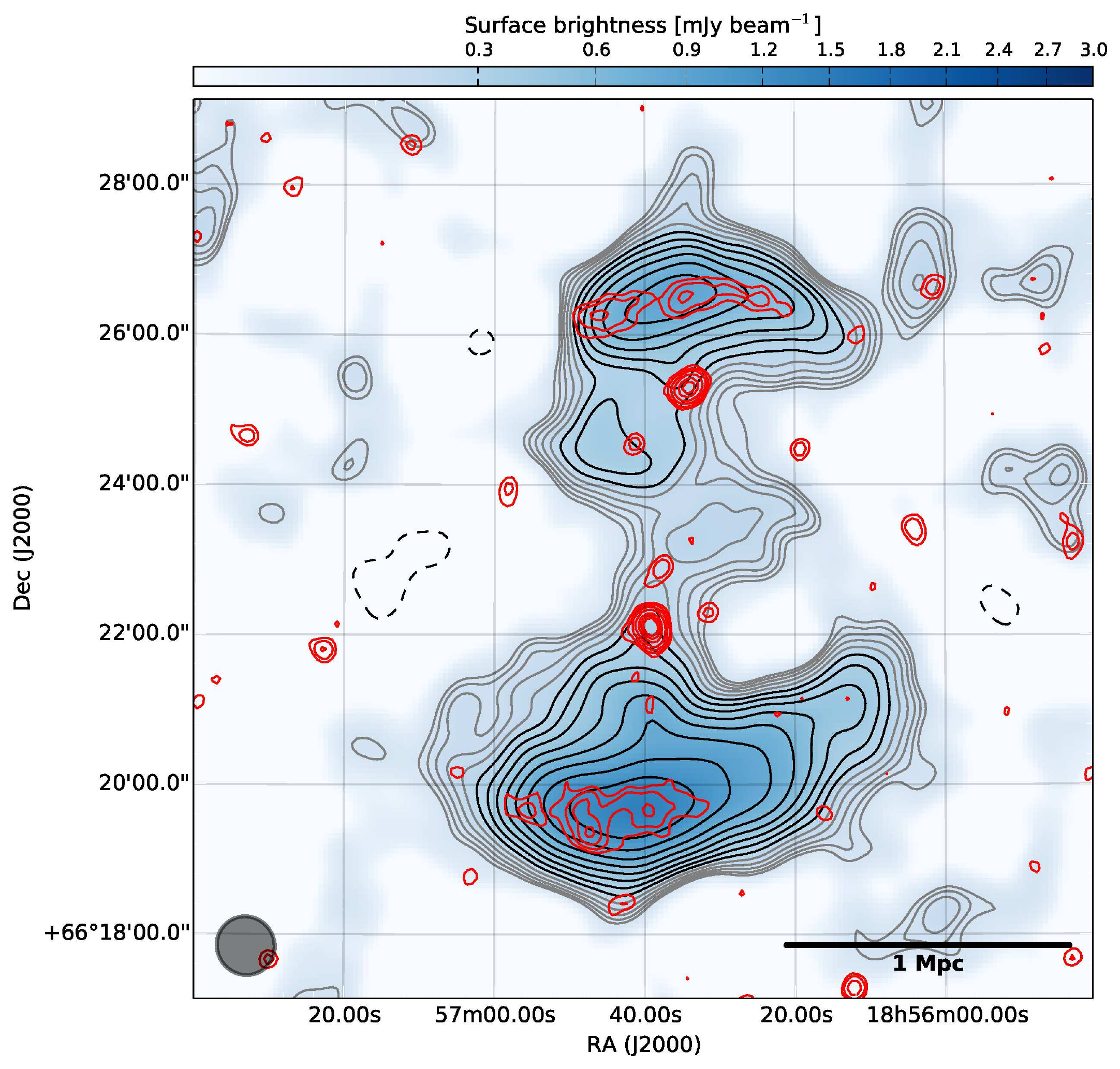}
\caption{WSRT 1.4 GHz image of the extended radio structure present in \target{}. The high-resolution image (red contours) has been subtracted and the residuals were imaged tapering the data to a resolution of 46\arcsec. Contours are logarithmically spaced between 1 and 20 $\sigma$, with $\sigma=90$~\mujybeam. Grey contours should be taken with caution since they are below $3\sigma$. Dashed contours are at $-3\sigma$.}\label{fig:halo}
\end{figure}

Finally, we detect traces of an extended, low-surface brightness feature extending from the two relics towards the centre of the cluster (see Fig.~\ref{fig:halo}). This can be a hint of a halo which could extend from the cluster centre up to the two radio relics, as in the system MACS J1752.0+4440 \citep{VanWeeren2012b}. Planned lower-frequency observations will confirm its presence. Similar to the case of the bridge in the Coma cluster \citep{Brown2011a}, the radio halo appears to be located in the vicinity of another radio source, implying a possible link.

\section{A catalogue of double radio relics}
\label{sec:catalogue}

\begin{table*}
\centering
\begin{tabular}{lccccccccccc}
\hline
Name                & $z$    & Freq   & Flux         & $-\alpha$ & $L_{1.4 \rm GHz}$       & LLS & $d_c$ & $d_r$ 
& & $L_{\rm X\ [0.1-2.4]}$   & $M_{500}$ \bigstrut[t]\\
                    &        & (GHz)  & (mJy)        &           & (10$^{23}$ W Hz$^{-1}$) &(kpc)& (kpc) & (kpc) 
& & (10$^{44}$ erg s$^{-1}$) & ($10^{14}$ \Msun) \bigstrut[b]\\
A3365 E             & 0.0926 & 1.425  & $42.6\pm2.6$ & --        & 7.4  & 547  & 984   & 1531
& \multirow{2}{*}{$\bigg\}$} & \multirow{2}{*}{0.86} & \multirow{2}{*}{--} \\
A3365 W             & 0.0926 & 1.425  & $5.3\pm0.5$  & --        & 0.9  & 291  & 1021  & 1531 & & & \\
ZwCl0008.8+5215 E   & 0.1032 & 1.382  & $56\pm3.5$   & 1.59      & 15.4 & 1416 & 944   & 1585
& \multirow{2}{*}{$\bigg\}$} & \multirow{2}{*}{0.5} & \multirow{2}{*}{$3.30^{+0.46}_{-0.50}$} \\
ZwCl0008.8+5215 W   & 0.1032 & 1.382  & $11\pm1.2$   & 1.49      & 3.0  & 303  & 640   & 1585 & & & \\
A2345 E             & 0.1765 & 1.425  & $29\pm0.4$   & 1.3       & 26.6 & 1565 & 1304  & 2348       
& \multirow{2}{*}{$\bigg\}$} & \multirow{2}{*}{4.3} & \multirow{2}{*}{$5.71^{+0.46}_{-0.49}$} \\
A2345 W             & 0.1765 & 1.425  & $30\pm0.5$   & 1.5       & 28.5 & 834  & 1096  & 2348 & & & \\
A1240 N             & 0.1590 & 1.425  & $6\pm0.2$    & 1.2       & 4.3  & 866  & 762   & 1802
& \multirow{2}{*}{$\bigg\}$} & \multirow{2}{*}{1.0} & \multirow{2}{*}{$3.71^{+0.50}_{-0.54}$} \\
A1240 S             & 0.1590 & 1.425  & $10.1\pm0.4$ & 1.3       & 7.3  & 1282 & 1039  & 1802 & & & \\
RXCJ1314.4-2515 E   & 0.2439 & 0.61   & $28\pm1.4$   & 1.41      & 16.8 & 1132 & 1023  & 1425
& \multirow{2}{*}{$\bigg\}$} & \multirow{2}{*}{10.9} & \multirow{2}{*}{$6.15^{+0.69}_{-0.73}$} \\
RXCJ1314.4-2515 W   & 0.2439 & 0.61   & $64.8\pm3.2$ & 1.40      & 39.0 & 913  & 621   & 1425 & & & \\
MACSJ1149.5+2223 E  & 0.544  & 1.450  & $4.1\pm0.2$  & 1.15      & 53.3 & 775  & 1476  & 2067
& \multirow{2}{*}{$\bigg\}$} & \multirow{2}{*}{14.0} & \multirow{2}{*}{$8.55^{+0.77}_{-0.82}$} \\
MACSJ1149.5+2223 W  & 0.544  & 1.450  & $5.6\pm0.3$  & 0.75      & 60.3 & 701  & 1108  & 2067 & & & \\
MACSJ1752.0+4440 NE & 0.366  & 1.714  & $55.1\pm2.9$ & 1.16      & 331  & 1190 & 1091  & 1917
& \multirow{2}{*}{$\bigg\}$} & \multirow{2}{*}{8.2} & \multirow{2}{*}{$6.96^{+0.53}_{-0.56}$} \\
MACSJ1752.0+4440 SW & 0.366  & 1.714  & $25.7\pm1.4$ & 1.10      & 150  & 694  & 892   & 1917 & & & \\
0217+70 E           & 0.065  & --     & --           & --        & --   & 1056 & 1021  & 1972
& \multirow{2}{*}{$\bigg\}$} & \multirow{2}{*}{0.63} & \multirow{2}{*}{--} \\
0217+70 W           & 0.065  & --     & --           & --        & --   & 1409 & 951   & 1972 & & & \\
A3376 E             & 0.046  & 1.4    & $122\pm10$   & 1.82      & 6.1  & 1180 & 638   & 1824
& \multirow{2}{*}{$\bigg\}$} & \multirow{2}{*}{2.12} & \multirow{2}{*}{$2.27^{+0.20}_{-0.21}$} \\
A3376 W             & 0.046  & 1.4    & $113\pm10$   & 1.70      & 5.7  & 1039 & 1149  & 1824 & & & \\
A3667 E             & 0.056  & 1.4    & $300\pm20$   & --        & 19.6 & 1430 & 1344  & 3347
& \multirow{2}{*}{$\bigg\}$} & \multirow{2}{*}{10.9} & \multirow{2}{*}{$5.77^{+0.24}_{-0.24}$} \\
A3667 W             & 0.056  & 1.4    & $3700\pm300$ & 1.1       & 271  & 2000 & 2000  & 3347 & & & \\
CIZAJ2242.8+5301 N  & 0.192  & 1.4    & $144\pm15$   & 1.06      & 150  & 1998 & 1585  & 2576
& \multirow{2}{*}{$\bigg\}$} & \multirow{2}{*}{6.8} & \multirow{2}{*}{--} \\
CIZAJ2242.8+5301 S  & 0.192  & 1.4    & $18\pm2$     & 1.29      & 19.5 & 1648 & 1175  & 2576 & & & \\
ZwCl2341.1+0000 N   & 0.27   & 0.6    & $14\pm3$     & 0.49      & 18.2 & 574  & 738   & 1907
& \multirow{2}{*}{$\bigg\}$} & \multirow{2}{*}{2.4} & \multirow{2}{*}{$5.15^{+0.64}_{-0.69}$} \\
ZwCl2341.1+0000 S   & 0.27   & 0.6    & $37\pm13$    & 0.76      & 40.9 & 1230 & 1189  & 1907 & & & \\
ACT-CLJ0102-4915 NW & 0.87   & 0.61   & $19\pm2$     & 1.19      & 296  & 500  & 1237  & 1590
& \multirow{2}{*}{$\bigg\}$} & \multirow{2}{*}{35} & \multirow{2}{*}{$8.80^{+0.64}_{-0.67}$} \\
ACT-CLJ0102-4915 SE & 0.87   & 0.61   & $3\pm0.3$    & 1.4       & 44.8 & 294  & 382   & 1590 & & & \\
PSZ1G287.0+32.9 NW  & 0.39   & 0.61   & $110\pm11$   & 1.16      & 233  & 2482 & 600   & 3135
& \multirow{2}{*}{$\bigg\}$} & \multirow{2}{*}{17.2} & \multirow{2}{*}{$13.89^{+0.53}_{-0.54}$} \\
PSZ1G287.0+32.9 SE  & 0.39   & 0.61   & $50\pm5$     & 1.33      & 97.1 & 1578 & 2700  & 3135 & & & \\
PSZ1G096.89+24.17 N & 0.3    & 1.381  & $8.9\pm0.8$  & --        & 15.2 & 880  & 770   & 1812
& \multirow{2}{*}{$\bigg\}$} & \multirow{2}{*}{3.7} & \multirow{2}{*}{$4.40^{+0.45}_{-0.48}$} \\
PSZ1G096.89+24.17 S & 0.3    & 1.381  & $18.3\pm1.9$ & --        & 31.2 & 1419 & 1145  & 1812 & & & \bigstrut[b]\\
\hline
\end{tabular}
\caption{Details of known double radio relics. Col. 1: source name. Col. 2: source redshift. Col. 3: frequency used for the radio flux measure. Col. 4: Radio flux. Col. 5: spectral index (when multiple spectral indices are available, we used the one describing the slope below 1.4 GHz), $S_\nu \propto \nu^{\alpha}$. Col. 6: k-corrected radio power. Col. 7: Largest linear scale of the relic. Col. 8: distance between the relic and the peak of the X-ray brightness. Col. 9: distance between the two relics. Col. 10: X-ray luminosity. Col. 11: cluster mass measured by the SZ effect.\label{tab:double_relic}}

A3365: \cite{VanWeeren2011a};
ZwCl0008.8+5215: \cite{VanWeeren2011};
A2345: \cite{Bonafede2009a, Boschin2010};
A1240: \cite{Bonafede2009a};
RXCJ1314.4-2515: \cite{Venturi2007a};
MACSJ1149.5+2223: \cite{Bonafede2012a};
MACSJ1752.0+4440: \cite{VanWeeren2012b, Bonafede2012a};
0217+70: \cite{Brown2011};
A3376: \cite{Kale2012a};
A3667: \cite{Rottgering1997, Johnston-Hollitt2004};
CIZAJ2242.8+5301 (``Sausage''): \cite{VanWeeren2010a, Stroe2013};
ZwCl2341.1+0000: \cite{Giovannini2010, VanWeeren2009a, Bagchi2011};
ACT-CLJ0102-4915 (``El Gordo''): \cite{Menanteau2012, Lindner2013a};
PSZ1G287.0+32.9: \cite{Bagchi2011a, Bonafede2014a};
PSZ1G096.89+24.17: this work
\end{table*}

To make a comprehensive study of double radio relic systems, we searched the literature for all double radio relics discovered so far. In table~\ref{tab:double_relic} we collected the data of 15 systems, for a total of 30 radio relics. All the lengths and powers have been recalculated using the cosmology of this paper. The luminosities have been k-corrected and scaled to a frequency of 1.4 GHz. In cases that no spectral index was available, we assumed a value of $\alpha=-1.3$ \cite{Feretti2012}. X-ray luminosities were also scaled to provide powers in the energy range $0.1-2.4$ keV. We also collected all the available mass estimations obtained by new SZ measurements \citep{PlanckCollaboration2013}. Abell 1758N could host a double radio relic \citep{Giovannini2009} but is not included in our catalogue since higher-resolution images are needed to confirm the presence of the relics.

For Abell 3365, 0217+70 and CIZA J2242.8+5301 the Planck catalogue does not provide a mass estimate. 
We used the Malmquist-bias corrected scaling relation provided by \cite{Pratt2009} which gives a mass estimate for a certain X-ray luminosity in the 0.1--2.4 keV range. For the CIZA J2242.8+5301 cluster we obtained $M_{500} = 7.97\pm0.35\times10^{14}\ \Msun$ while for Abell 3365 we obtained $M_{500} = 2.01\pm0.08\times10^{14}\ \Msun$. Since no radio fluxes are available in the literature for 0217+70, we exclude this system from the subsequent analysis.

\subsection{Revising old correlations}

\cite{VanWeeren2009b} collected the available data in the literature searching for correlations in the radio relics properties. They found that the relic LLS correlates with the projected distance from the cluster centre, with larger relics located at larger distances from the cluster centre, and that the spectral index anti-correlates with the relic LLS, with smaller relics having a steeper spectral indices. These correlations can be explained by the fact that the larger shock waves occur mainly in lower density and lower temperature regions, and hence have larger Mach numbers \citep{Skillman2008,Vazza2009}. 

However, as already recognized by the authors, all length measurements are affected by projection effects. Furthermore, radio phoenices\footnote{We consider radio phoenices those emissions generated by fossil radio plasma produced by an old episode of AGN activity. The plasma is compressed, by e.g. a merger shock wave, which boosts both the magnetic field inside the plasma and the momenta of the relativistic particles.} could contaminate the sample adding small-scale, steep spectrum objects that are typically close to the cluster centre. 

To overcome these problems, \cite{Bonafede2012a} selected only double radio relics, which in the most plausible scenario are consequences of mergers seen in the plane of the sky. This would remove the presence of radio phoenices and minimizes the impact of the projection effect on the measured lengths. They found a good correlation only for the LLS--$L_{1.4 \rm GHz}$ relation, meaning that relics with high radio power tend to have larger linear sizes.

Here we have repeated their analysis including six new radio relics (three double relic systems), re-measuring all lengths and distances to be consistent with our cosmology and adding the cluster mass taken from the SZ measurements. Linear regression has been made using the BCES bisector method described in \cite{Akritas1996a} which accounts for errors in, both, the dependent and independent variable and for the intrinsic scatter of the data.

To access the existence of possible correlations between the selected quantities, we use Spearman's rank correlation coefficient (S-rank). The null hypotheses we are testing is the following: `there is no association between variable A and variable B'.

\begin{figure*}
\centering
\subfloat[]{\includegraphics[width=\columnwidth]{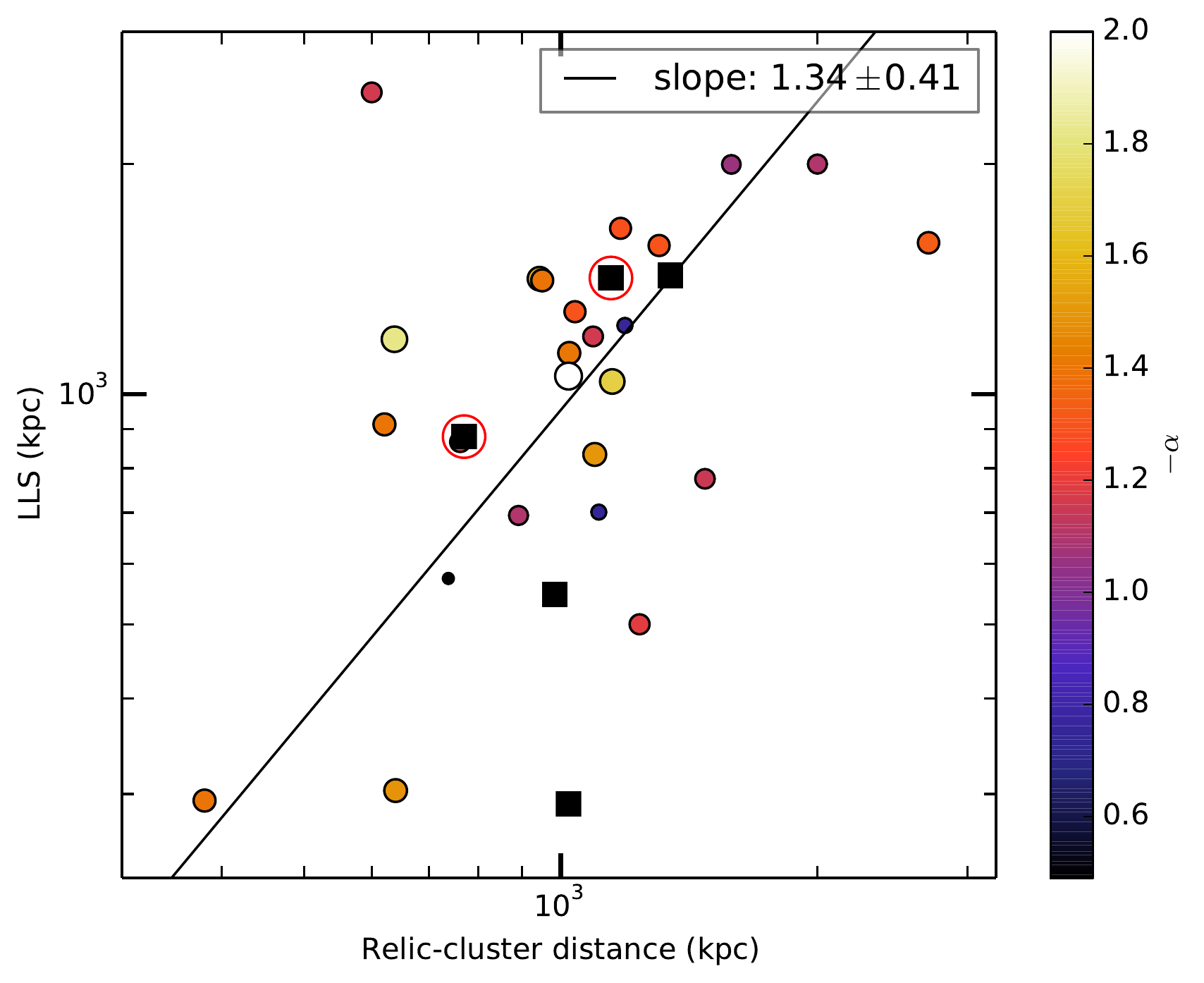}\label{fig:cdist-LLS}}
\subfloat[]{\includegraphics[width=\columnwidth]{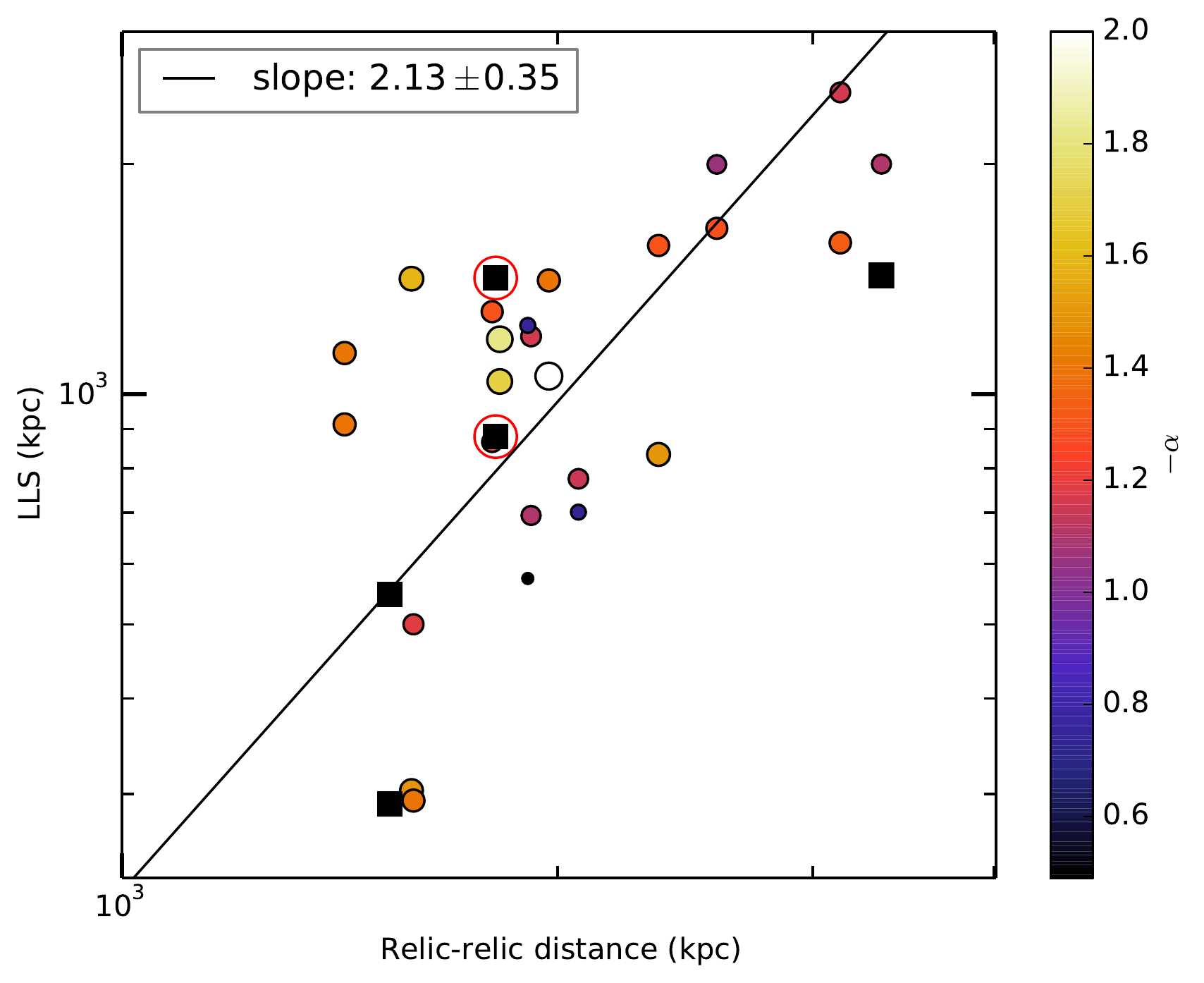}\label{fig:rdist-LLS}}
\caption{Left panel: LLS versus the projected distance between the radio relics in double systems and the X-ray peak of the hosting cluster. Color-coded is the spectral index value, black squares are for unknown values of the spectral index. The solid line shows the linear regression. Red circles show the values for \target. Right panel: same as the left panel but the independent variable is now the projected distance between the two relics of each system.}\label{fig:dist-LLS}
\end{figure*}

The null hypothesis is accepted with a significance level of 83\% for the spectral index -- LLS correlation (meaning that no correlation is present) as also found by \cite{Bonafede2012a}. The same correlation seen in \cite{VanWeeren2009b} was most likely driven by the presence of steep-spectrum, centrally located radio phoenices which are excluded from our sample. The null hypothesis is instead rejected (meaning correlation) with a 2\% significance level for the LLS correlation with the cluster distance (see Fig.~\ref{fig:dist-LLS}). Although we have recalculated all distances between radio relics and the cluster centre positions assuming the peak of the X-ray emission, this can be misleading since all systems are undergoing a merger which makes it difficult to define a proper cluster centre. Hence, we have recalculated this correlation using the distance between the two radio relics as the independent variable. In this case, we find a better correlation (null hypothesis rejected at $0.03$\% significance level, see Fig.~\ref{fig:dist-LLS}), which supports the idea that the shock surfaces are larger the further out they are. However, we do not find a significant correlation in the $L_r$ -- LLS plane as found by \cite{Bonafede2012a}.  

\subsection{Mass-luminosity correlation}

\begin{figure}
\centering
\subfloat[Double radio relic systems]{\includegraphics[width=\columnwidth]{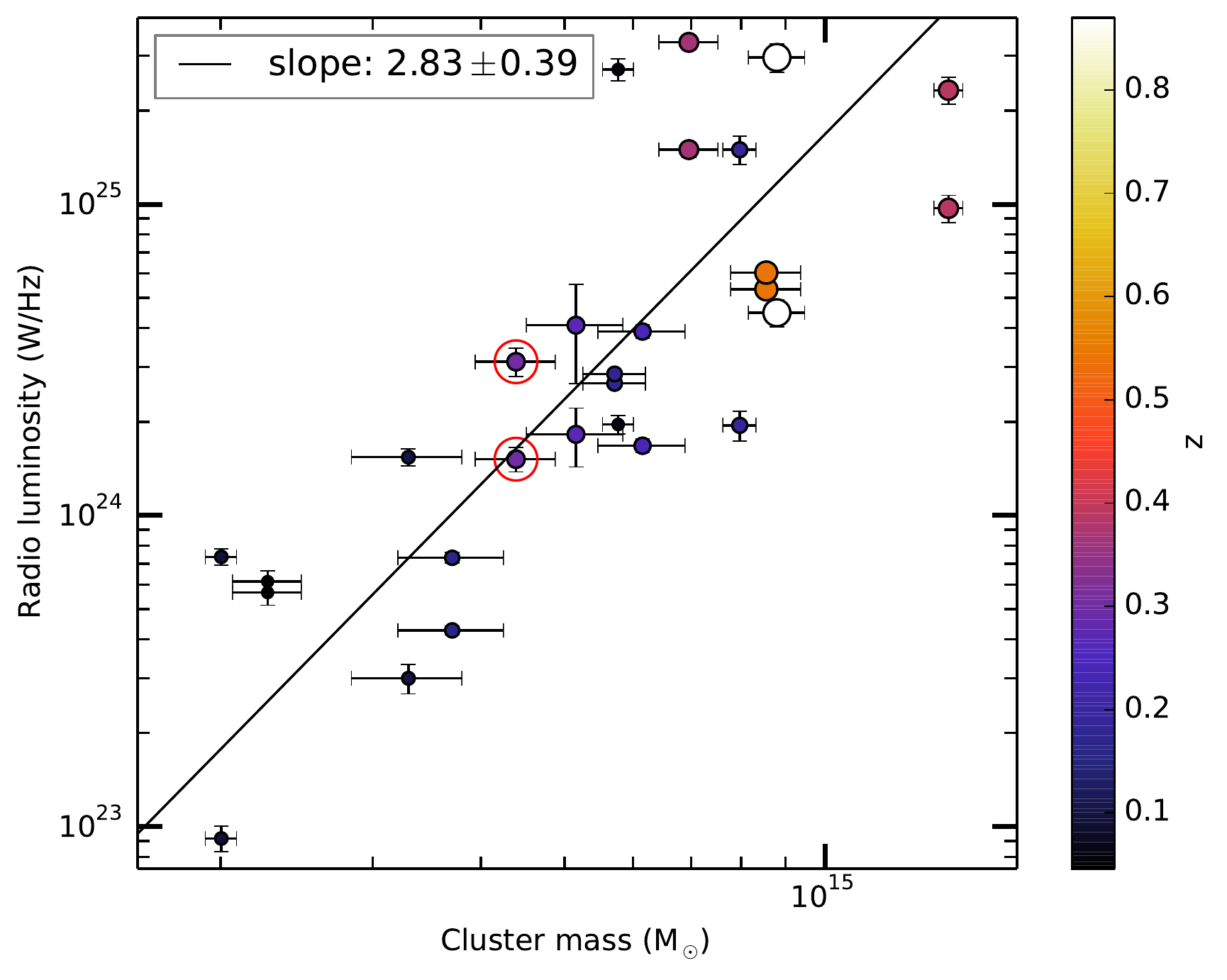}\label{fig:dr_Mass-Lr}}\\
\subfloat[Double + single radio relic systems]{\includegraphics[width=\columnwidth]{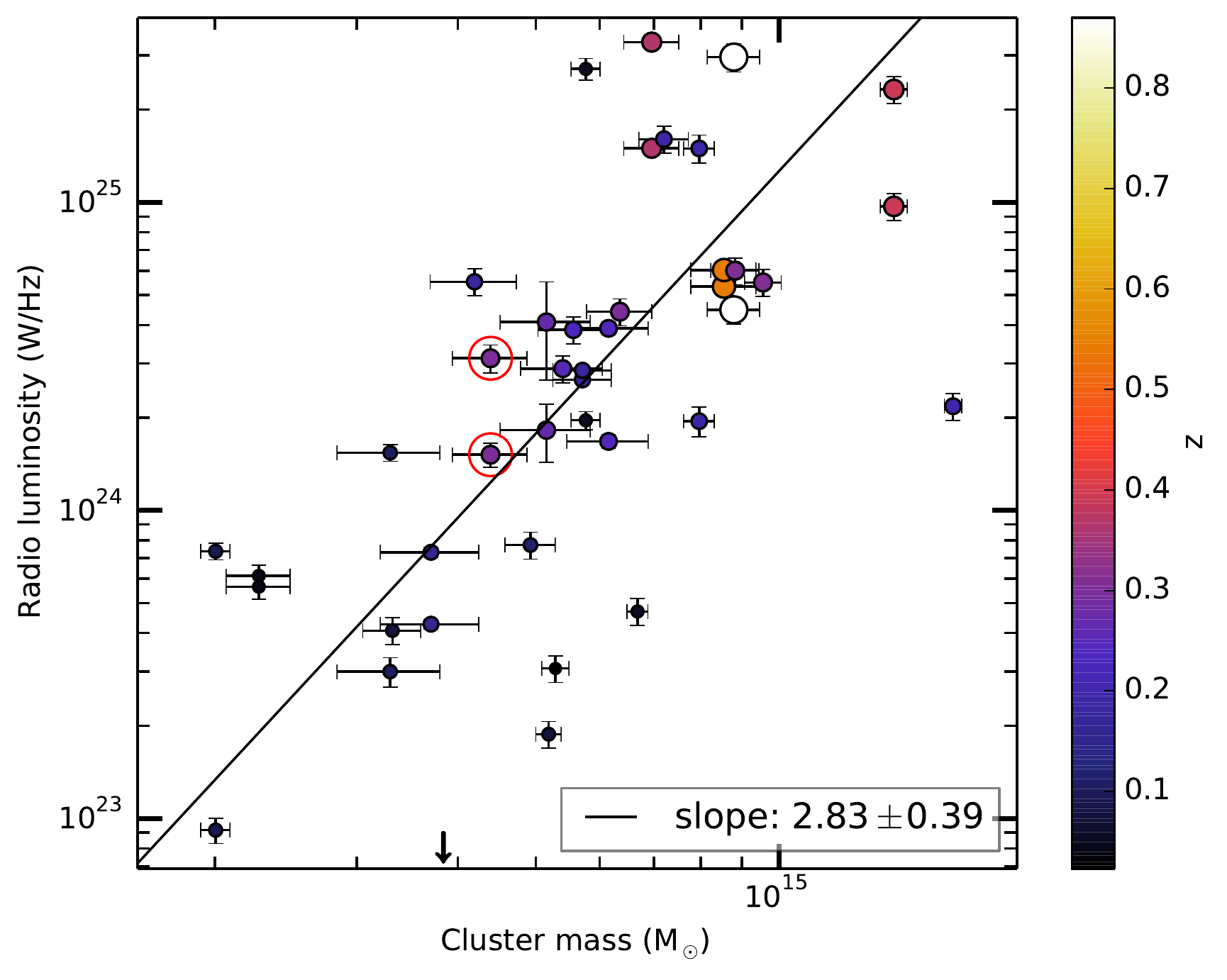}\label{fig:drsr_Mass-Lr}}
\caption{Cluster mass against radio relic power. Color coded the cluster redshift. Solid line is the linear regression. Red circles show the values for \target. Top: for double radio relics presented in table~\ref{tab:double_relic}. Bottom: for double radio relics plus all relics of \citet{Feretti2012} marked as ``extended'' for which the mass is available from Planck SZ measurements. The arrow indicates the peculiar case of Abell 2146 where a clear shock is detected in the X-ray without a radio counterpart \citep{Russell2011}.}\label{fig:Mass-Lr}
\end{figure}

We have found a remarkably good correlation between the luminosity of the double radio relics and the mass of the host clusters (see Fig.~\ref{fig:Mass-Lr}). To test this relation further, we have included also single radio relics from \cite{Feretti2012}. We excluded those relics classified by the authors as ``roundish'', which are likely to be radio phoenices, and those relics for whom no Planck SZ mass is given. The error on the radio fluxes for these relics are assumed to be 10\% of the total flux. The correlation in this case is less strong but holds with the null hypothesis rejected at $<0.1$\% significance level (see Fig.~\ref{fig:Mass-Lr}). In both cases, the slope of the regression is $\sim 2.8\pm0.4$.

Both samples plotted in Fig.~\ref{fig:Mass-Lr} are heterogeneous and affected by several observational and selection biases. All the low-mass ($M_{500}<5\times10^{14}$~\Msun) clusters are relatively close-by objects ($z\lesssim0.35$), although they would be an order of magnitude more abundant in the redshift range $0.35<z<0.9$ than in the local universe due to the larger sampling volume \citep[see e.g.,][]{Vikhlinin2009}. This effect is likely driven by a selection bias due to difficulties in finding high-redshift low-mass clusters. On the other other hand, the high-mass end of the sample should have a similar number of high-redshift ($z>0.35$) and low-redshift ($z<0.35$) objects, given a lesser impact of the selection bias \citep{PlanckCollaboration2013}. We find indeed four double radio relic systems for the high and five for the low redshift bin.

\begin{figure}
\centering
\includegraphics[width=\columnwidth]{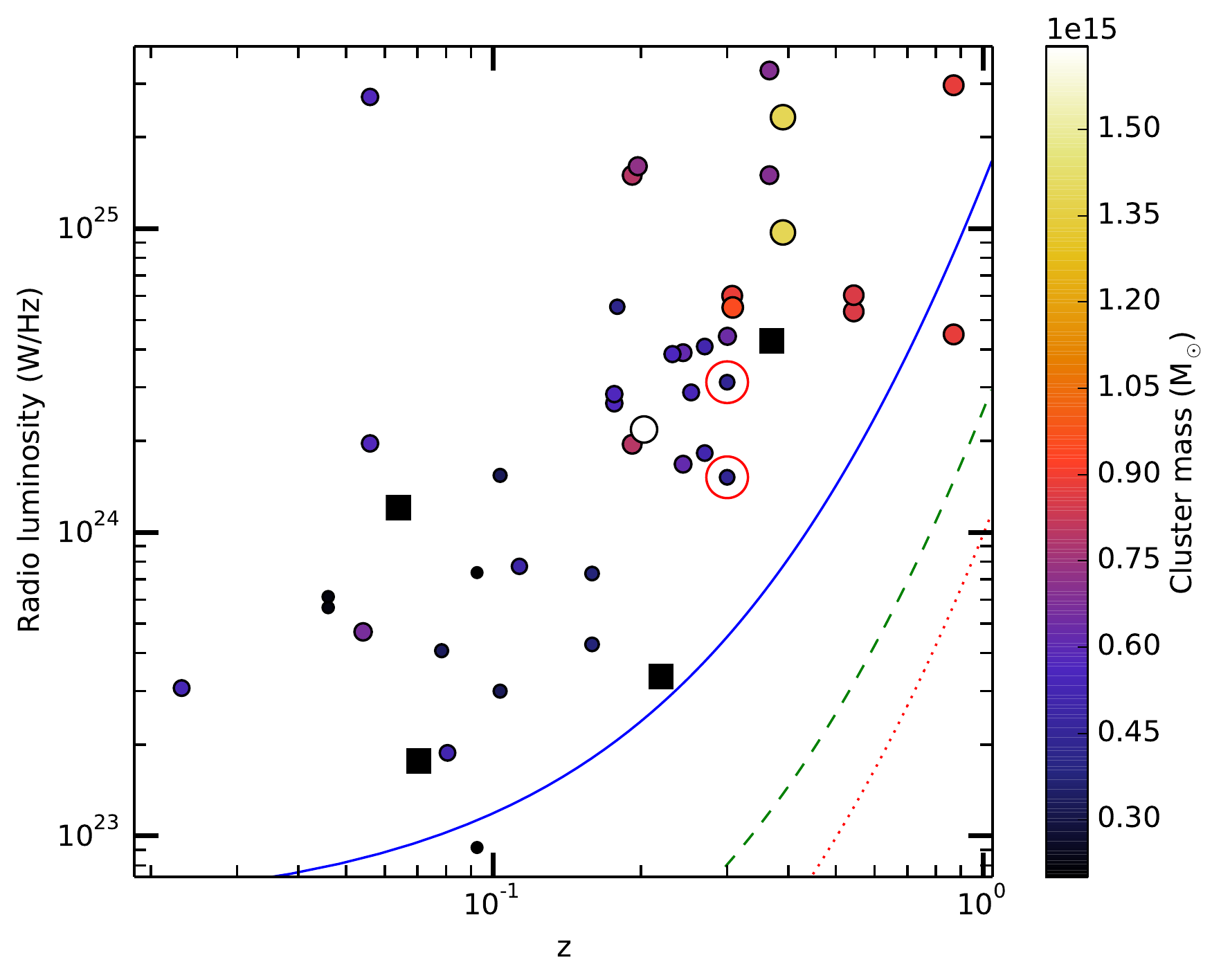}
\caption{Cluster redshift versus radio relic luminosity for all double radio relics listed in Table~\ref{tab:double_relic} and all single relics presented in \citet{Feretti2012} marked as ``extended''. Black squares are clusters for which the Planck SZ catalogue does not provide a mass estimate. Solid blue line is an estimated $1\sigma$ detection limit for relic detection in NVSS, green dashed line is the confusion limit for the VLA in D-configuration (which can be reached in 1 minute observation with the EVLA), while red dotted line is for LOFAR Tier-1 (rms 0.1 mJy and resolution \beam{25}{25}). Red circles show the values for \target.}\label{fig:Lr-z}
\end{figure}

The absence of high-power radio relics in low-mass clusters (top-left corner of Fig.~\ref{fig:Mass-Lr}) comes from observational evidence: low-massive clusters are unable to produce high-luminosity radio relics. However, the absence of low-power radio relics in massive clusters (bottom-right corner of Fig.~\ref{fig:Mass-Lr}) can be an observational bias driven by the distance dependency of the two variables (Malmquist bias). The effect is explained in Fig.~\ref{fig:Lr-z} where we plot the cluster redshift versus the radio luminosity of radio relics in our sample. The lines show an estimated $1\sigma$ detection limit for some typical observation set-ups assuming a relic size of $1000\times100$ kpc$^{2}$ and taking cosmic dimming into account. This plot shows why no low-luminosity relics are detected at high-redshift. However, all massive clusters ($M_{500}>5\times10^{14}$~\Msun) at moderate redshift lie well above the sensitivity cuts and the known close-by ($z<0.25$), massive clusters (RXCJ1314.4-2515, CIZAJ2242.8+5301 and Abell 3667) are the only ones at low-redshift that host high-luminosity radio relics. These two facts could give some confidence that the relation is not caused by an observation bias. However, higher sensitivity surveys, as those planned with LOFAR, are required to address this issue satisfactorily.

\subsection{Clues from simulations}

If we assume that radio relics are caused by merger shocks and that the luminosity of radio relics is proportional to the total energy released in a cluster merger, we can attempt to find a theoretical justification for the observed relation.

Simulations \citep[e.g.,][]{Poole2006} suggest that a fixed fraction of the total energy dissipated in a merger goes into the primary merger shocks. The energy released in a merger between clusters of mass $M$ is given by $E \propto M^2/R_{\rm vir}$, where $R_{\rm vir}$ is the virial radius that is proportional to $R_{\rm vir} \propto M^{1/3}$. If the observed radio power scales as the dissipated kinetic energy flux, then the luminosity of the relics is $P\propto E/t_{\rm cross}$, where $t_{\rm cross}$ is the sound crossing time of the cluster (this is an assumption for the lifetime of the relics). The sound crossing time is $R_{\rm vir}/c_{\rm s}$, where $c_{\rm s}$ is the speed of sound which is $\propto T^{1/2}$. According to the mass-temperature relation \citep[e.g.,][]{Pratt2009}, $T\propto M^{2/3}$ so that
$t_{\rm cross} \propto M^0 = {\rm const}$. Hence, this simple argument would predict a relation of the kind $P \propto M^{5/3}$. This is shallower than the slope of 2.8 found above. The discrepancy could be caused by the dependence of the particle acceleration efficiencies or the magnetic fields at the relics on some variable related to the cluster mass. Alternatively, this could be the result of observational bias, or the life-time of the relic depends on the cluster mass. Possibly, more insights may be gained from simulations.

The relation between the power of radio relics and the host cluster mass and luminosity has been investigated with cosmological simulations in the past \citep[e.g.][]{Hoeft2008,Pfrommer2008,Skillman2011}. However, previous work has been based on very small samples, often has made no clear distinction between radio relics and radio halos, and has been limited by numerical resolution.

Here we compare to a sample of clusters simulated with the AMR hydrodynamics code ENZO \citep{Bryan2014} by \citet{Vazza2010}, which consists of 20 galaxy clusters with masses in the range of $6 \times 10^{14} \leq M/M_{\odot} \leq 3 \times 10^{15}$ extracted from a total volume of $L_{\rm box} \approx 480$ Mpc/h, and probing down to a resolution of $\approx 25 ~\rm kpc/h$ for most of the virial volume. We analysed this cluster sample at redshifts $z=0$, $z=0.3$ and $z=0.6$, and extracted the Mach number and kinetic energy flux across shocks measured in the 3D distribution of gas (as in \citealt{Vazza2012}). Then we assigned a radio power to each shocked cell based on the model of \cite{Hoeft2007}. To do so, we explored different models for the distribution of magnetic fields: a) uniform fields with $B=1\ \mu\rm G$ everywhere ({\it Buni} model); b) the magnetic field model derived by \cite{Bonafede2010} as a best fit for the magnetic field profile of the Coma cluster ({\it Bderiv} model); c) we assigned to each shocked cell in the simulation a magnetic field corresponding to a 10\% equipartition with the thermal gas energy, $B_{\rm eq}=\sqrt{0.1\ E_g \cdot 8 \pi}$, where $E_g$ is the thermal gas energy of each cell ({\it Bequip} model). This way, the (maximum) local amplification of the magnetic field is related to the thermal energy density of the ICM and to the host cluster mass. For each object and redshift we generated mock radio observations, and analysed them with a 2D relic detection scheme, which identifies as part of the same relic all pixels brighter than a given threshold as in \citet{Vazza2012}. This way we can select objects with a clear presence of large double relic systems and compare them to the sample of observed relics.

\begin{figure}
\centering
\label{fig:sim}
\includegraphics[width=\columnwidth]{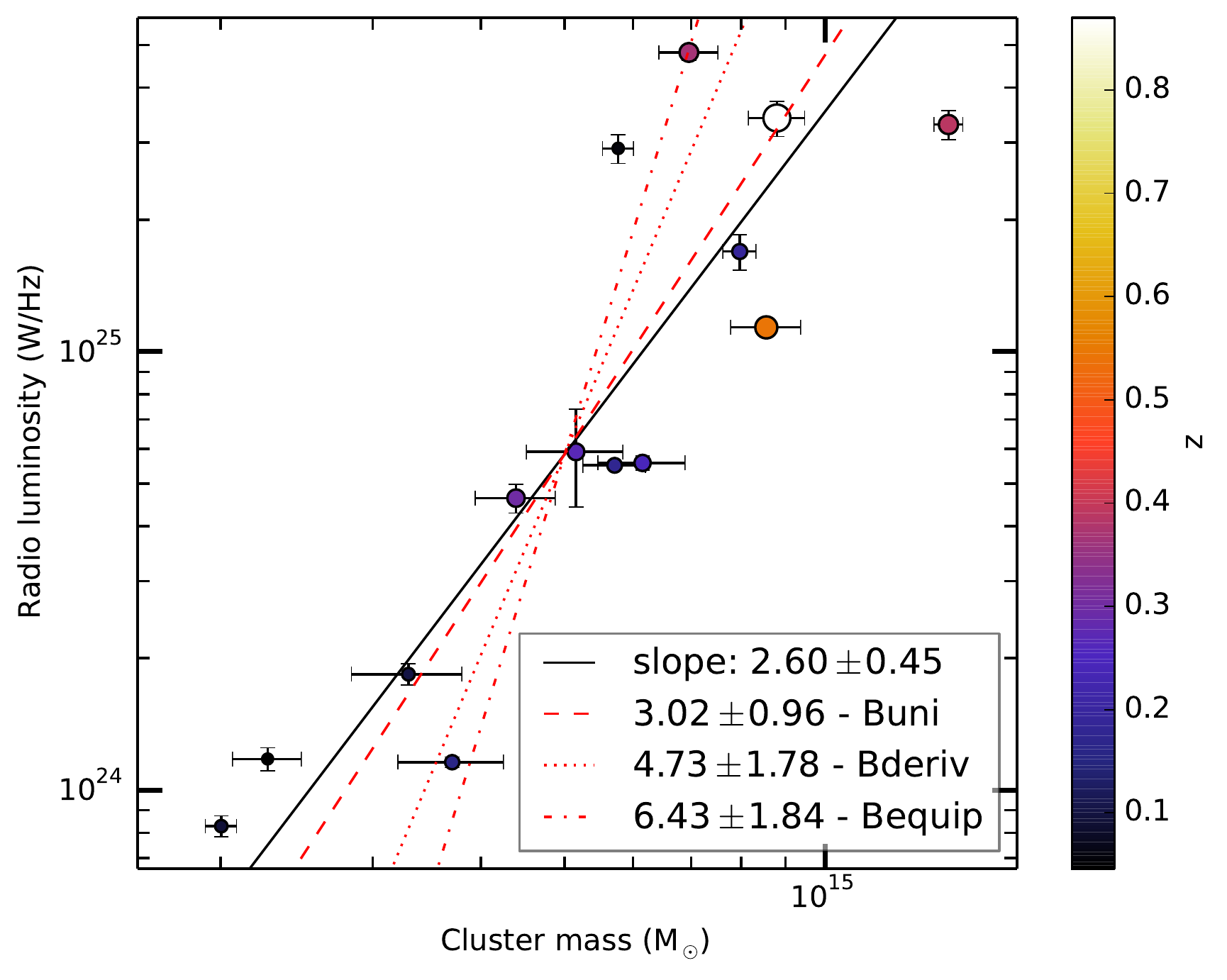}
\caption{Same as in Fig.~\ref{fig:dr_Mass-Lr} but the power of radio relics belonging to the same clusters is added up. Red dashed line show the relation as predicted by simulations for a uniform magnetic field. Dotted line is using the magnetic field derived by the best fit of the Coma cluster data \citep{Bonafede2010}. Dash-dotted line is 10\% of equipartition. Model lines were rescaled to match the data linear fit at $M=5\times10^{14}$~\Msun}\label{fig:coup_M-Lr-z}
\end{figure}

Fig.~\ref{fig:coup_M-Lr-z} shows the comparison between the total radio power of the observed systems together with the best-fit of simulated relics for each model. Using the total radio emission instead of the power of the two radio relics separately is more robust against possible numerical uncertainties related to the specific criterion used to excise the radio power of each relic object in the simulated image. Furthermore, it is motivated by the scenario in which the total energy dissipated by each couple of relics follow the same merger event. While the normalization of the relation is dependent on the values of the magnetic field and of the electron acceleration efficiency \citep[e.g.,][]{Hoeft2007}, the slope is not. Therefore, we renormalized all simulated best-fit relations in Fig.~\ref{fig:coup_M-Lr-z} to match the observations best-fit at $M_{500} = 5 \times 10^{14} M_{\odot}$. The figure shows that while the fixed magnetic field model is compatible with the observed scaling ($L_{1.4 \rm GHz} \propto M^{3.0 \pm 1.0}$), the other two models yield steeper relations ($4.73\pm1.74$ for the {\it Bderiv} model and $6.43\pm1.84$ for the {\it Bequip} model), incompatible with the observed relation. So it may be that the magnetic field does not vary much across the clusters in the sample and a strength of $\sim 2\ \mu\rm G$ provides a reasonable fit to the observed relation. 

The simulated ($L_{\rm r}$, $M$) relation is steeper than the analytical estimate previously given ($L_r \propto M^{5/3}$), even assuming a fixed magnetic field strength. The likely reason is that simulated radio relics always follow cluster-cluster mergers, where significant departures from the virial scaling previously presented are observed. In particular, simulated mergers present an abrupt increase of X-ray emission and temperature \citep[e.g.,][]{Poole2006}, due to the fast dissipation of kinetic energy in merger shocks. The same process should lead to an even sharper increase of the radio power in relics \citep{Skillman2013}, whose dependence on cluster mass is more complex than derived from virial relations. However, we note that the normalisation is also changed by assuming a different acceleration efficiency for the electrons (which is of order of $\sim 10^{-4}$ here). As a cautionary remark, one should note that recent works of the radio and $\gamma$-ray properties of clusters that have double relics \citep{Vazza2014} has cast doubts on the validity of the standard shock acceleration scenario that we just explored with this set of simulations.

\section{Conclusions}
\label{sec:conclusions}

We have reported the discovery of a double radio relic system in the galaxy cluster \target. We present WSRT observations of the cluster at 1.4 GHz which show two giant radio relics with a linear size of 0.9 Mpc and 1.4 Mpc and a hint of the presence of a radio halo. The short distance from the cluster centre and the rotated polarization angle for the southern relic suggest that the merger axis is slightly tilted compared to the plane of the sky, with the southern relic farther away from the observer.

We have compiled a comprehensive catalogue of all the known double radio relics in the literature. Compared to single radio relics, double systems are rarely misclassified radio phoenices and their projected distance are likely close to the real one. Using this catalogue we revisited the scaling relations investigated by \cite{VanWeeren2009b} and \cite{Bonafede2012a}. Compared with previous works, we found no correlation between the LLS and the spectral index, but we confirm the correlation between the LLS and the relic distance from the cluster centre, in line with the idea that in the periphery of the clusters the shock surfaces are larger.

Our catalogue was then cross-matched with the cluster masses provided by the SZ measurements made by Planck. We found a good correlation between the radio relic luminosities and the cluster masses ($L_r \propto M^{2.83\pm0.39}$). The relation could be reproduced in cosmological simulations and suggests that the magnetic field at the location of radio relics is rather uniform in all relics (within a factor of a few, which would explain the observed scatter around the mean), even though their masses, radial distances and Mach numbers are different.

\section*{Acknowledgements}
The authors thank Gianfranco Brunetti for the useful discussions. AB and MB acknowledge support by the research group FOR 1254 funded by the Deutsche Forschungsgemeinschaft: "Magnetisation of interstellar and intergalactic media:the prospects of low-frequency radio observations".

\bibliographystyle{mn2e}
\bibliography{PSZ1G096.89+24.17}
\bsp

\label{lastpage}

\end{document}